\documentclass[letterpaper,twocolumn,11pt]{article}

\newlength\titlebox
\usepackage{times}
\usepackage{helvet}
\usepackage{courier}
\usepackage{url}
\usepackage{graphicx}
\usepackage{booktabs}
\usepackage{multirow,rotating}
\usepackage[usenames,dvipsnames]{color}
\usepackage{amsmath}

\usepackage[]{authblk}  

\usepackage[font=small]{caption}
\usepackage{abstract}
\date{March 28, 2012}
\usepackage[compact]{titlesec}

\newcommand{\mat}[1]{\mathbf{#1}}
\newcommand{\bmat}[1]{\boldsymbol{#1}}
\usepackage{enumitem}

\def \t0{$t_0$}
\def \tn{$t^*$}
\def \te{$t_e$}

\def \rt1{$\mathbf{rt}_{t-1}$}
\def \rp1{$\mathbf{rp}_{t-1}$}
\def \src1{$\mathbf{src}^{\alpha}_{t-1}$}
\def \fol1{$\mathbf{follow}^{\alpha}_{t-1}$}
\def \rtE1{$\mathbf{rtEnv}_{t-1}$}
\def \rpE1{$\mathbf{rpEnv}_{t-1}$}
\def \srcE1{$\mathbf{srcEnv}^{\alpha}_{t-1}$}

\def \rtT{$\mathbf{rt}^{\alpha}_{t^*}$}
\def \rpT{$\mathbf{rp}^{\alpha}_{t^*}$}
\def \srcT{$\mathbf{src}^{\alpha}_{t^*}$}
\def \folT{$\mathbf{follow}^{\alpha}_{t^*}$}
\def \rtEt{$\mathbf{rtEnv}_{t}$}
\def \rpEt{$\mathbf{rpEnv}_{t}$}
\def \srcEt{$\mathbf{srcEnv}^{\alpha}_{t}$}

\newcommand{\specialcell}[2][l]{%
  \begin{tabular}[#1]{@{}l@{}}#2\end{tabular}}

\def\dm#1{{\small\color{Blue}\textbf{[DM: #1]}}}
\def\bk#1{{\small\color{Magenta}\textbf{[BK: #1]}}}

\def\bk#1{}
\def\dm#1{}

\def\figfilepath{figs}
\graphicspath{{\figfilepath/}}

\title{\#Bigbirds Never Die: Understanding Social Dynamics of \protect\\ Emergent Hashtags}

\author[1,*]{Yu-Ru Lin}
\author[2]{Drew~Margolin}
\author[1]{Brian Keegan}
\author[3]{Andrea Baronchelli}
\author[1,2]{David Lazer}
\affil[1]{College of Social Sciences and Humanities, Northeastern University, Boston, MA 02115, USA}  
\affil[2]{College of Computer and Information Science, Northeastern University, Boston, MA 02115, USA}
\affil[3]{Lab for the Modeling of Biological and Socio-technical Systems, Northeastern University, Boston, MA 02115, USA}  
\affil[*]{To whom correspondence should be addressed. Email: \protect\url{yuruliny@gmail.com}}

\begin{document} 

\twocolumn[
    \maketitle
    \vspace{-3em}
    \begin{onecolabstract}
        \vspace{-1em}
We examine the growth, survival, and context of 256 novel hashtags during the 2012 U.S. presidential debates. Our analysis reveals the trajectories of hashtag use fall into two distinct classes: ``winners'' that emerge more quickly and are sustained for longer periods of time than other ``also-rans'' hashtags. We propose a ``conversational vibrancy'' framework to capture dynamics of hashtags based on their topicality, interactivity, diversity, and prominence. Statistical analyses of the growth and persistence of hashtags reveal novel relationships between features of this framework and the relative success of hashtags. Specifically, retweets always contribute to faster hashtag adoption, replies extend the life of ``winners'' while having no effect on ``also-rans.'' This is the first study on the lifecycle of hashtag adoption and use in response to purely exogenous shocks. We draw on theories of uses and gratification, organizational ecology, and language evolution to discuss these findings and their implications for understanding social influence and collective action in social media more generally.
    \end{onecolabstract}
]

\section{Introduction}

The hashtag is a ubiquitous and flexible annotation for Twitter users. Hashtags allow users to track ongoing conversations, signal membership in a community, or communicate non-verbal cues like irony. Hashtags vary considerably in their activity and audience, but little is known about how the use of novel hashtags co-evolve with the needs of their users and the presence of other hashtags. Why do some hashtags persist while others are just momentary blips? How do similar hashtags compete for attention?

Hashtags often reflect eccentric topics and their emergence is happenstance. These idiosyncrasies typically limit researchers' ability to systematically compare features of their emergence and evolution. However, the shared attention to and improvisational nature of the U.S. presidential debates provides a unique opportunity to understand the social dynamics of novel hashtag adoption following repeated exogenous events. Before the October 2012 debates, the importance of concepts such as ``big bird'', ``malarky'', ``binders'', or ``bayonets'' to a political campaign would be minimal. Yet these unexpected terms now denote salient moments from these debates and became rallying calls for partisans and punchlines for comedians. More broadly, Twitter users created hundreds of hashtags related to these topics which had no \textit{raison d'\^etre} before the debates. The candidates' unscripted statements can thus be understood as creating an exogenous shock to the system of political discourse.  Combined with fine-grained data about large scale and real time user behavior create, these exogenous shocks create a set of natural experiments to systematically analyze the features that contribute to the growth and stabilization of hashtags.

Previous research has examined the roles of ``relevance'' and ``exposure'' for hashtag adoption~\cite{huang2010conversational,yang2012we,lehmann2012dynamical,yang2010counts,romero2011social,romero2011differences}. However, these features are inadequate for distinguishing hashtag trajectories in the context of an event as prominent as the U.S. presidential debates where relevance and exposure are effectively ``maxed out.'' Furthermore, hashtags are more than labels for contextualizing statements, objects for bookmarking, or channels for sharing information, but they are active virtual sites for constructing communities. We propose an alternative framework called \textit{conversational vibrancy} to understand the dynamics of novel hashtag use based on how features such as topicality, interactivity, diversity, and prominence interact with the communities producing and consuming tweets to these hashtags.

We operationalize our conversational vibrancy framework with behavioral and structural variables such as retweets, replies, unique tweets, and audience size. Analysis of the relationship between these features and the rate and sustainability of hashtag adoption reveal the existence of two distinct classes of hashtag trajectories. ``Winners'' are hashtags characterized by more rapid growth and longer sustained interest than ``also-rans.''  Statistical models of the growth and persistence of each class of hashtags reveals novel and complex interactions with features from our conversational vibrancy framework as well as the concentration of user attention. Our analysis suggests conversational vibrancy contributes to the growth and sustenance of hashtags in different phases of their life-cycles. For example, while retweets and followers support a hashtag's growth, they also paradoxically undermine its persistence. We draw on theories of organizational ecology and language evolution to discuss these findings and their implications for understanding social influence and collective action in social media more generally.


\section{Background}
Twitter is a popular ``micro-blogging'' platform founded in 2006 that allows users to receive and broadcast (up to) 140-character messages called ``tweets''. The brevity and immediacy of the medium contributed to its rapid adoption as a way of sharing information from mobile devices, propagating information by ``retweeting'', as well as tracking relevant content by labeling tweets with ``hashtags''~\cite{boyd2010tweet}. 

Hashtags first emerged on Twitter during the 2007 San Diego wildfires as a way to track relevant information about a large scale natural disaster by labeling content so that it could be filtered and shared via backchannels~\cite{sutton2008backchannels}. Unlike traditional tagging systems used for information archival, Twitter hashtags can serve either as a label for identifying topically relevant streams of message or a prompt for commenting and sharing~\cite{huang2010conversational}. Hashtags often fill a dual role as both a topical identifier (\textit{e.g.}, \#iPhone) and a symbol of a community membership (\textit{e.g.}, \#VoteForObama)~\cite{yang2012we}. During the ``Arab Spring'' and other protests, hashtags were used by activists to coordinate their actions and garner support~\cite{starbird2012will}.

Studies of hashtag use have primarily emphasized two dimensions: relevance and exposure. The relevance perspective emphasizes the dependence hashtags have on the topics being discussed: content relevant to the Twitter community draw more tweets and retweets and thus grow sustainably over time~\cite{huang2010conversational,yang2012we,lehmann2012dynamical,yang2010counts}. From the view of exposure, however, hashtag's publicity and legitimacy is driven by social exposure as part of a process of complex contagion: hashtags grow because they are adopted by users with social connections to one another, and by users' repeated use and spreading the word to others of interests~\cite{romero2011social,romero2011differences}. 

Analyses of hashtag adoption have characterized differences in the persistence of hashtags, modeled the spread of hashtags as a type of complex contagion requiring multiple exposures, and the behavior of hashtags that emerge in response to exogenous shocks~\cite{romero2011differences,romero2011social}. Other studies of hashtag have focused on the temporal occurrences of the hashtags that are characterized by either ``peaky'' but ephemeral topics versus ``persistent conversations'' using less salient terms over longer periods of time~\cite{shamma2011peaks}. In~\cite{lehmann2012dynamical}, peaks in popular hashtags are the result of four discrete classes of activity and are primarily driven by exogenous rather than endogenous factors. While these analyses have analyzed the use of hashtags in the context of specific events, topics, or communities, they have not examined the ecology of multiple hashtags competing for attention following an exogenous event. 

\subsection{Conversational vibrancy}
While prior work has emphasized relevance and exposure as mediators of hashtag activity, these features are inadequate for explaining activity during events such as elections, sports, and awards shows that attract unusual levels of shared attention and intense activity. The large audiences who view these popular events predispose them to generating hashtags that become widely adopted as users ``dual screen'' by tweeting about what they watch~\cite{shamma2009tweet,shamma2011peaks}. During these events, exogenous shocks can simultaneously expose users to a large number of candidate hashtags that are of similar relevance.  These constructs are thus inadequate to distinguish between hashtags that emerge under these conditions. A more interesting question is why are some hashtags related to these events adopted while others are not? 

Uses and gratification (U\&G) is classic theoretical paradigm from mass communication research that examines why people become involved in some types of mediated communication versus others and what gratifications they receive from it. U\&G emphasizes ``what people do with media rather than what media do to people'' which makes it especially salient for studying social media like Twitter where users both produce and consume content. U\&G argues that different forms of media compete to provide distinct uses such as news or entertainment and audiences actively seek out those media that are meaningful to them and can best satisfy their individual needs for information, social connection, or emotional support~\cite{ruggiero2000uses,stafford2004determining}. 

We draw from U\&G theory to argue that users choose to employ some hashtags while ignoring others because patterns of use around these hashtags provide a better match for their needs. However, we also recognize that patterns of  communication~\cite{boyd2010tweet} and perceptions of users' audiences~\cite{marwick2010boyd} interact in complex ways with other contextual and content-level features~\cite{yang2010counts,suh2010hong} to influence the spread and adoption of content in social media~\cite{lerman2010ghosh}. This suggests the need for an integrative framework that synthesizes these multilevel processes to understand why users adopt some hashtags while rejecting others. We propose \textit{conversational vibrancy} as an explanatory framework to understand the emergence, growth, and persistence of a hashtag and its community of users. Conversational vibrancy consists of four general elements, each of which we operationalize with behavioral features specific to Twitter:
\begin{itemize}[itemsep=2pt,parsep=0pt,topsep=0pt, partopsep=0pt]
\item \textit{Topicality} is a kind of relevance that reflects the extent to which tweets labeled by a hashtag are timely and have contextual relevance for on-going conversations. ``\#bigbird'' was highly topical following the first debate but was not topical during the inauguration because its humor and relevance are lost in this different context. The number of times a hashtagged tweet created by one user is retweeted reflects the value this information has for other users. Hashtags are more likely to grow and persist when users re-share hashtagged content created by others with their own networks.
\item \textit{Interactivity} is the extent to which individuals attend to others' statements by responding to them in turn. Hashtags can denote spaces for on-going interpersonal conversation, supportive interest-based communities, or heated debates that bring others into the community and get them to adopt the hashtags as well. A hashtag may encourage people to launch discussions about the topic or people may simply reply to a user with a hashtag to acknowledge his or her wit. The number of replies co-occurring with a hashtag captures the extent to which users are not only paying attention to others' content but actively responding to one another. Hashtags are more likely to grow and persist when there are many users speaking directly to each other while invoking the hashtag.
\item \textit{Diversity} captures whether activity is concentrated in the same information or people are attending to many different sources of information. Hashtags that emerge because of a single ``one hit wonder'' tweet being retweeted over and over are qualitatively different from many unique tweets using the same hashtag. Because counting the absolute frequency of tweets observed using a hashtag can be biased towards hashtags being hustled by spammers, the number of unique tweets that have been retweeted at least once provides a reasonable threshold for capturing the breadth of tweets attracting attention. Hashtags are more likely to grow and persist when many unique tweets with the hashtag are being shared.
\item \textit{Prominence} measures the audience exposed to a hashtag by a user mentioning it. The audience of Twitter users plays a crucial and complex role in information sharing behaviors~\cite{marwick2010boyd}. Features of users themselves sharing information can have greater influence than the content itself being shared \cite{yang2010counts}. Hashtags mentioned by users with large audiences of many followers are more likely to result in rapid growth and sustained activity than hashtags mentioned by users with small audiences.
\end{itemize}
In summary, hashtags exhibiting higher levels of conversational vibrancy should gratify more users' needs and thus contribute to the success of the hashtag. The user-generated and socially-mediated nature of Twitter will create a positive feedback loop through which hashtags with high conversational vibrancy will preferentially recruit more attention from users who in turn increase the vibrancy of the hashtag while also decreasing the communicative vibrancy of competing hashtags. We discuss these evolutionary dynamics and ecological interactions in more detail in the discussion.

\section{Our Approach}
\begin{table*}[!tb]
    \caption{Examples of debate-related tweets from the first debate on 4 October 2012}
    \label{tab:tweets}
    \centering\small 
    \begin{tabular}{ lcl }
    \toprule
hashtag&time (EDT)&tweet text\\
    \midrule
bigbird2012&21:28:12   &\specialcell{\#bigbird2012}\\
bigbird&21:28:12             &\specialcell{Romney stopping subsidy to PBS telling the host that?! wooo \#BigBird}\\
savebigbird&21:28:15     &\specialcell{\#savebigbird \#debate}\\
bigbird&21:28:18             &\specialcell{How you cut \#BigBird though \#Romney....Wtf??? Smh}\\
bigbird&21:28:20             &\specialcell[t]{@MittRomney says he will cut funding to PBS even thou he likes Lehrer and \#bigbird \#debates}\\
savebigbird&21:28:50     &\specialcell{Cut PBS? Noooooooooooi \#SaveBigBird}\\
supportbigbird&21:46:37&\specialcell[t]{RT @BlGBlRD: Yo Mitt Romney, Sesame Street is brought to you today by the letters F U!\\ \#debates \#SupportBigBird}\\
supportbigbird&21:51:54&\specialcell{This entire election is now about who will save Big Bird. \#supportbigbird \#debates}\\
supportbigbird&21:52:45&\specialcell{\#OccupySesameStreet \#SupportBigBird we are the 47\%}\\
    \bottomrule
    \end{tabular}
\vspace{-.2cm}
\end{table*}

Contemporary presidential campaigns in the United States are highly scripted and candidates offer repetitive ``stump speeches'' that rarely provide new information about their position or strategies~\cite{skewes2007}. However, debates are highly ceremonial and enthralling events that not only attract large audiences of pundits, partisans, and undecided voters, but also require candidates to respond directly to critiques and improvise responses outside of well-worn campaign talking points~\cite{schroeder2008}. As a result, these debates are ripe for unexpected and embarrassing statements. These ``gaffes'' can be impromptu exchanges that make a candidate look foolish or outright factual errors that undermine a candidate's credibility and often later become fodder for opponents, sometimes substantially changing the dynamics of a race ~\cite{clayman2006}. Examples of ``gaffes'' from the first presidential debate include Mitt Romney's unprompted threat to cut funding for the popular PBS children's television show ``Sesame Street'' with its iconic lead character Big Bird as well as Barack Obama's widely-derided aloof performance.

The combination of fine-grained behavioral data about a large population of people and high-stakes circumstances of Presidential debates provide a unique opportunity to analyze how a large and engaged audience reacts to statements to which it had not previously been exposed. Because substantial portions of the audience are monitoring or participating in social media while simultaneously watching the live debate~\cite{pew2012dualscreening}, these gaffes immediately prompt behavior on social media as users try to confirm a statement or alternatively improvise humorous responses to mock it. Unlike prior studies of hashtag adoption, threats to validity from confounded endogenous processes are minimized since gaffes and other unexpected statements made by candidates are inherently unpredictable and exogenous from our analysis. 

Examples of tweets containing the hashtags from the first debate are provided in Table~\ref{tab:tweets}. The table shows tweets that were posted immediately after or close to the onset time of different hashtags, focusing on the most popular, politically relevant hashtags or their variants. In the first debate, after Mitt Romney mentioned ``Big Bird,'' there were several similar hashtags, including \#bigbird2012, \#bigbird and \#savebigbird that appeared simultaneously in users' tweets (around 21:28 EDT). The \#supportbigbird hashtag was popularized 15 minutes after it, largely due to intensive retweeting of a newly-created account under the handle of ``@BIGBIRD''. 


We examine the dynamics of hashtag adoption during and immediately following each of the four debates during the 2012 presidential campaign. To understand why some novel hashtags are adopted while others are not, we measure two specific outcomes: the initial \textit{growth} of the hashtag as users adopt it and the \textit{persistence} of hashtag use after this initial growth phase. Because the behavior we model in this project is highly domain-specific to politics, we identify and preferentially sample from a population of politically active users to extract their tweeting history. For this sample of users, we examine the emergence of hashtags which are ``novel'' because they had not appeared in users' feeds beforehand and identify 256 of the most popular and domain-relevant hashtags. Using a variety of inferential statistical modeling approaches, we examine the relationship between the features of the conversational vibrancy framework and the growth and persistence of these hashtags.

\begin{table*}[!tb]
    \caption{Summary of datasets. All times are Eastern Daylight Time (EDT).}
    \label{tab:dataset}
    \centering\small 
    \begin{tabular}{ l llll }
    \toprule
Debate number & 1 & 2 (Vice Presidential) & 3 & 4\\
    \midrule
Debate starting time& 3 Oct. 21:00 & 11 Oct. 21:00 & 16 Oct. 21:00 &22 Oct. 21:00 \\
Tweet volume at peak&3,268,918&2,388,963&3,608,291&2,415,703\\
Unique users at peak&174,297&155,739&181,329&152,538\\
``Novel'' hashtags&92,432&58,165&91,705&77,526\\
``Pop'' hashtags&75&57&82&42\\
Tracking conclusion time & 7 Oct. 02:00&15 Oct. 02:00&20 Oct. 02:00&26 Oct. 02:00\\
    \bottomrule
    \end{tabular}
\vspace{-.1cm}
\end{table*}

\subsection{Dataset}

\begin{figure*}[!tb]
\hspace*{-.8cm}
\begin{tabular}{cccc}
(a) DEB 1 & (b) DEB 2 & (c) DEB 3 & (d) DEB 4\\
        \includegraphics[trim=.42cm 0cm .38cm 0cm, clip=true,width=.24\textwidth]{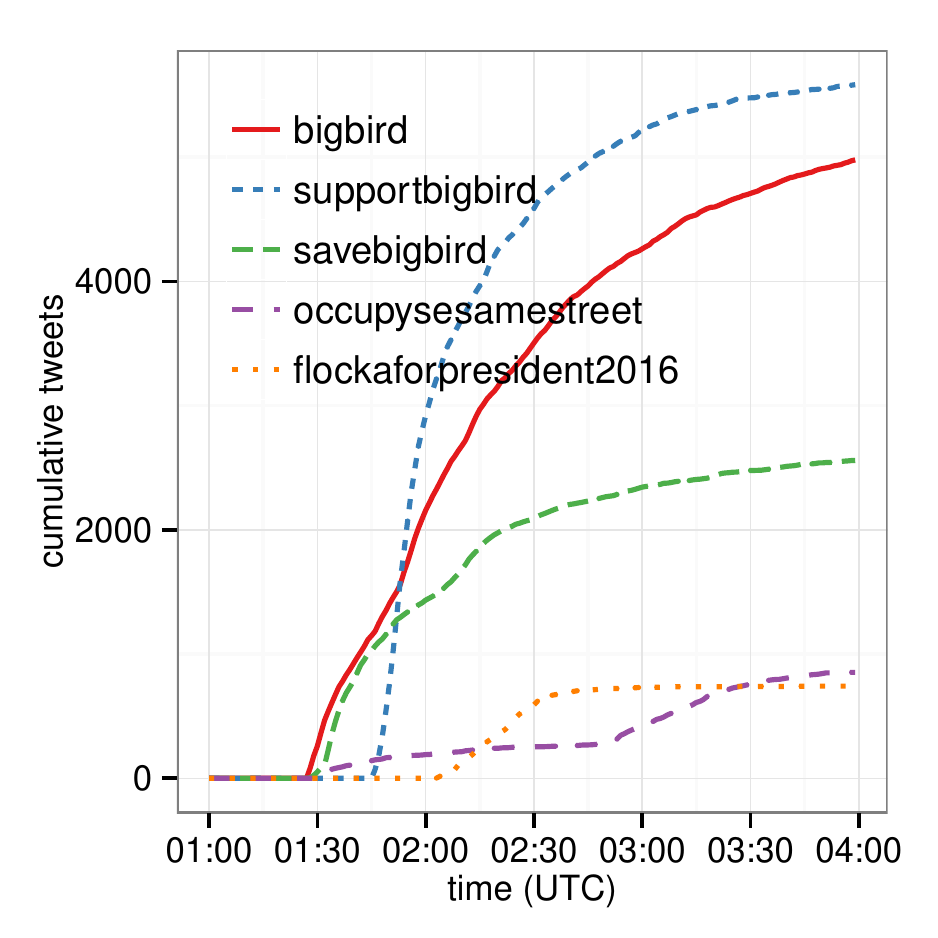} &
        \includegraphics[trim=.42cm 0cm .38cm 0cm, clip=true,width=.24\textwidth]{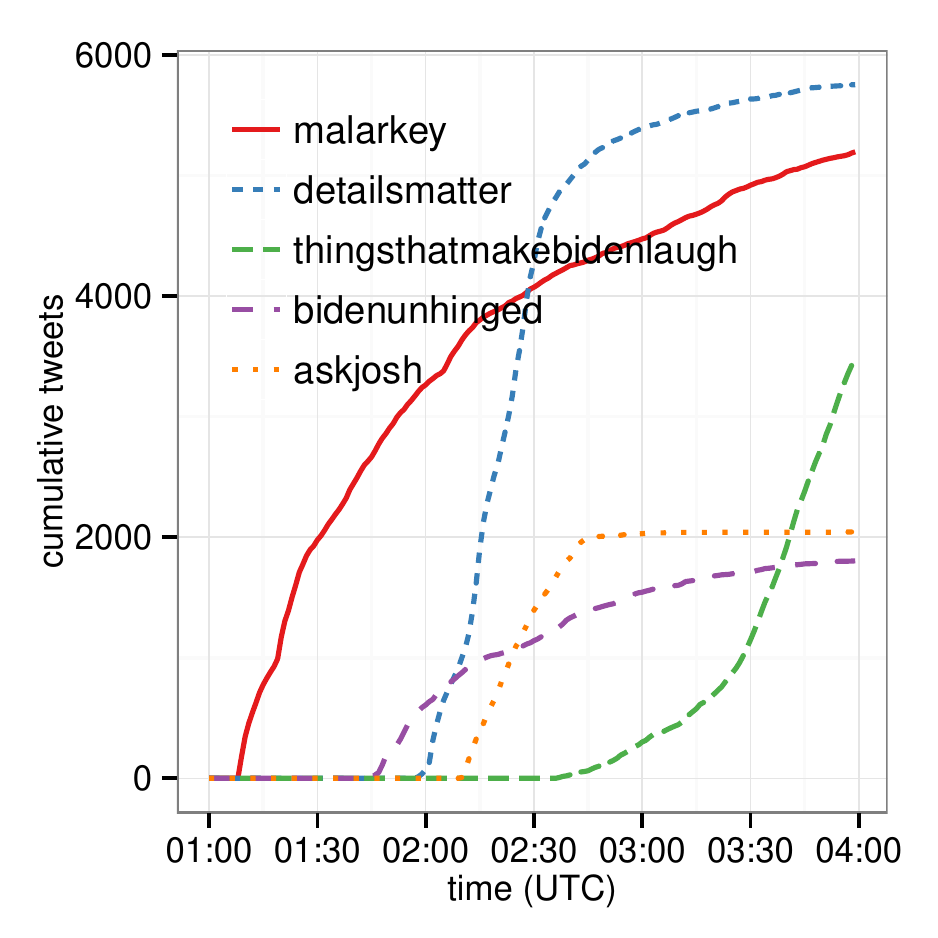} &
        \includegraphics[trim=.42cm 0cm .38cm 0cm, clip=true,width=.24\textwidth]{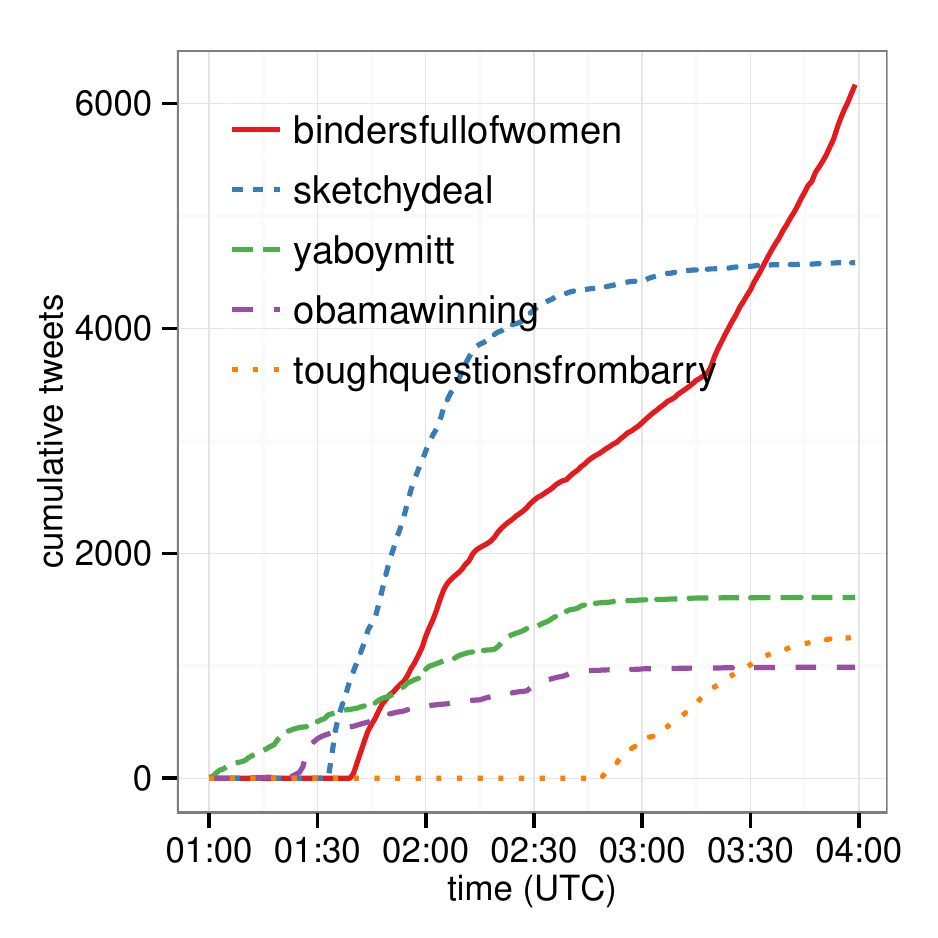} &
        \includegraphics[trim=.42cm 0cm .38cm 0cm, clip=true,width=.24\textwidth]{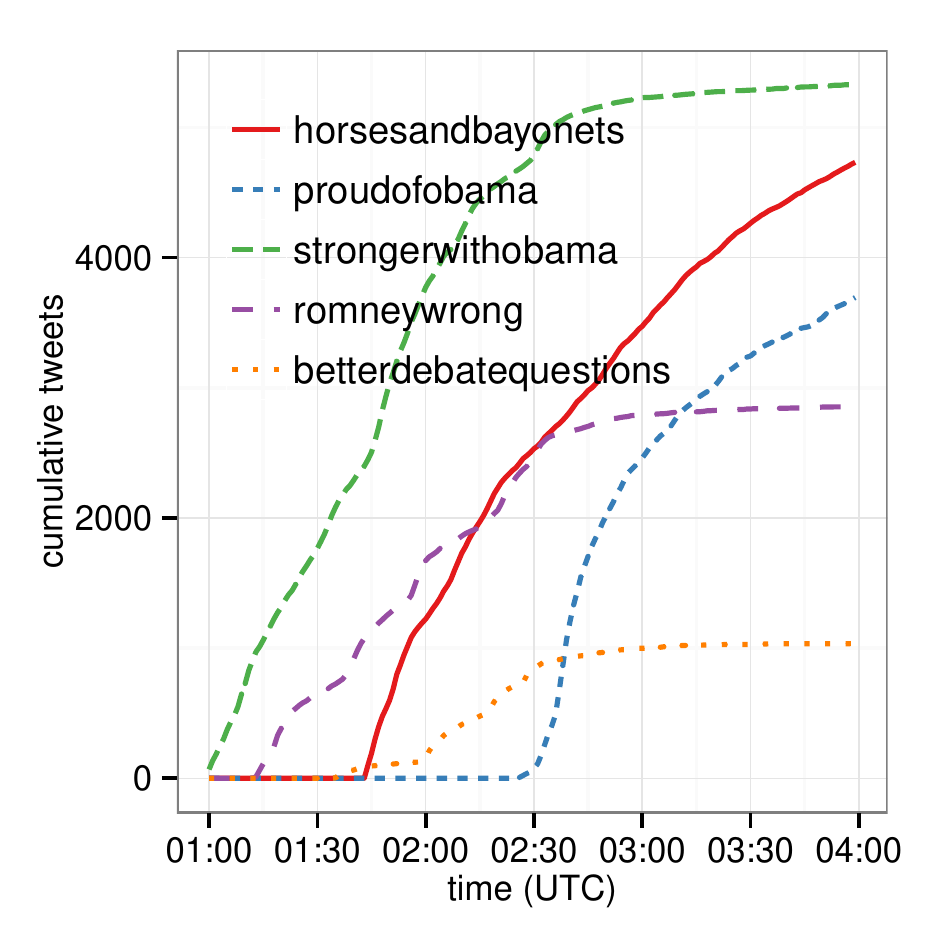}\\
\end{tabular}
\vspace{-2em}
    \caption{%
Cumulative tweet volume of top hashtags over time, starting from each debate. 
        \label{fig:debate_3hr}}
\vspace{-.6cm}
\end{figure*}

While randomly sampling content from Twitter's ``garden hose'' and aggregating it to understand collective behavior is methodologically defensible under some research designs, we argue it is inappropriate for our purposes. We identify a relevant sub-population of users as follows. Users with relevant characteristics are identified first, and tweets for these users are extracted for analysis.  This focus on the tweets from specified users reduces the threat of selection bias where inclusion in the observed set is correlated with the dependent variable~\cite{lin2012meter}.  The dependent variables in this study refer to changes in tweet volumes over time.  We thus focus on a fixed set of users---those that tweeted heavily during the debates---across the entire observation period.  Changes in tweets to the hashtag can thus be attributed to choices made by these users. 

\textbf{User sampling.} First, we identified politically-active users who tweeted using a hashtag such as ``\#debate'' or mentioned either candidate's Twitter account\footnote{@barackobama, @mittromney} during any of the four presidential debates. Using Twitter's ``garden hose'' streaming API\footnote{https://dev.twitter.com/docs/streaming-apis}, users whose tweets appeared in the feed were selected into our population. Second, we extracted these users' tweeting histories beginning in mid-August 2012 through late October 2012 using Twitter's REST API\footnote{https://dev.twitter.com/docs/api/1.1}. Because these queries are expensive owing to rate limits, we prioritized users who tweeted during more of the debates. Thus users who tweeted during all four debates are more likely to be represented in the sample than users who tweeted during only one of the debates. While biased, this population captures users such as journalists, pundits, politicians, and activists who are more politically active and thus relevant to our research setting. The subset of the resulting corpus used in this analysis contains 123,560,785 tweets from 2,516,125 unique users posted between September 29 and October 27.

Descriptive statistics for each of the four debates are summarized in Table~\ref{tab:dataset}. Each debate occurred between 21:00 and 22:30 EDT and we tracked the behavior of our population of users for 77 hours following each debate. The tweet volume peaked at the first hour of each debate and remained high at the second hour. We refer the two-hour window as ``peak window'' and identify ``novel'' hashtags that were born in each of the peak windows. Figure~\ref{fig:debate_3hr} shows the growth of top five hashtags in terms of the cumulative number of tweets for each debate.

\textbf{Pre-processing.} We extract ``novel'' hashtags as all hashtags that were contained in tweets posted during one of the peak windows but did not appear in any tweets posted within 96 hours prior to the peak windows. For each novel hashtag, we extract the users who have mentioned the hashtag within the 77 hour post-debate window starting from the peak window. We then select hashtags that have been mentioned by at least 100 unique users to be ``pop'' hashtags. Despite the features of our politically active user population, many of these hashtags were not related to the debate or politics generally and were removed unless the content of the tweet contained terms such as ``debate,'' ``president,'' or any of the four candidates' names. The resulting corpus contained 256 ``pop'' hashtags that were novel, widely used, and relevant to the debates. The numbers of pop hashtags from each debate are reported in Table~\ref{tab:dataset}.

The conversational vibrancy framework outlined above is operationalized using the following variables.
\begin{enumerate}[itemsep=2pt,parsep=0pt,topsep=0pt, partopsep=0pt]
\item \textit{Number of retweets} ($\mathbf{rt}$) under the hashtag is an indicator of its \textbf{topicality}. A retweet is an example of an individual identifying a tweet that is so interesting or relevant that they wish to re-state it, verbatim, to their own followers. Hashtags attracting more retweets have higher topicality.
\item \textit{Number of replies} ($\mathbf{rp}$) to a hashtag is an indicator of its \textbf{interactivity}. A heated debate may be more interactive than a collection of statements that agree but do not address one another.
\item \textit{Number of unique retweet sources} ($\mathbf{src}$) measures the \textbf{diversity} of the conversation. A hashtag with more unique retweet sources provides more diverse information. 
\item \textit{Expected follower size} ($\mathbf{follow}$) is proxy for the \textbf{prominence} of the users who tweet to the hashtag. Users with many followers tweet to a hashtag will  increase the growth and persistence of the hashtag.
\end{enumerate}

\subsection{Characterizing Hashtag Growth and Persistence}
\begin{figure}[!tb]
\hspace*{-.8cm}
\begin{tabular}{c}
(a) after 6 hours\\
        \includegraphics[trim=.38cm 0cm .38cm 0cm, clip=true,width=1\columnwidth]{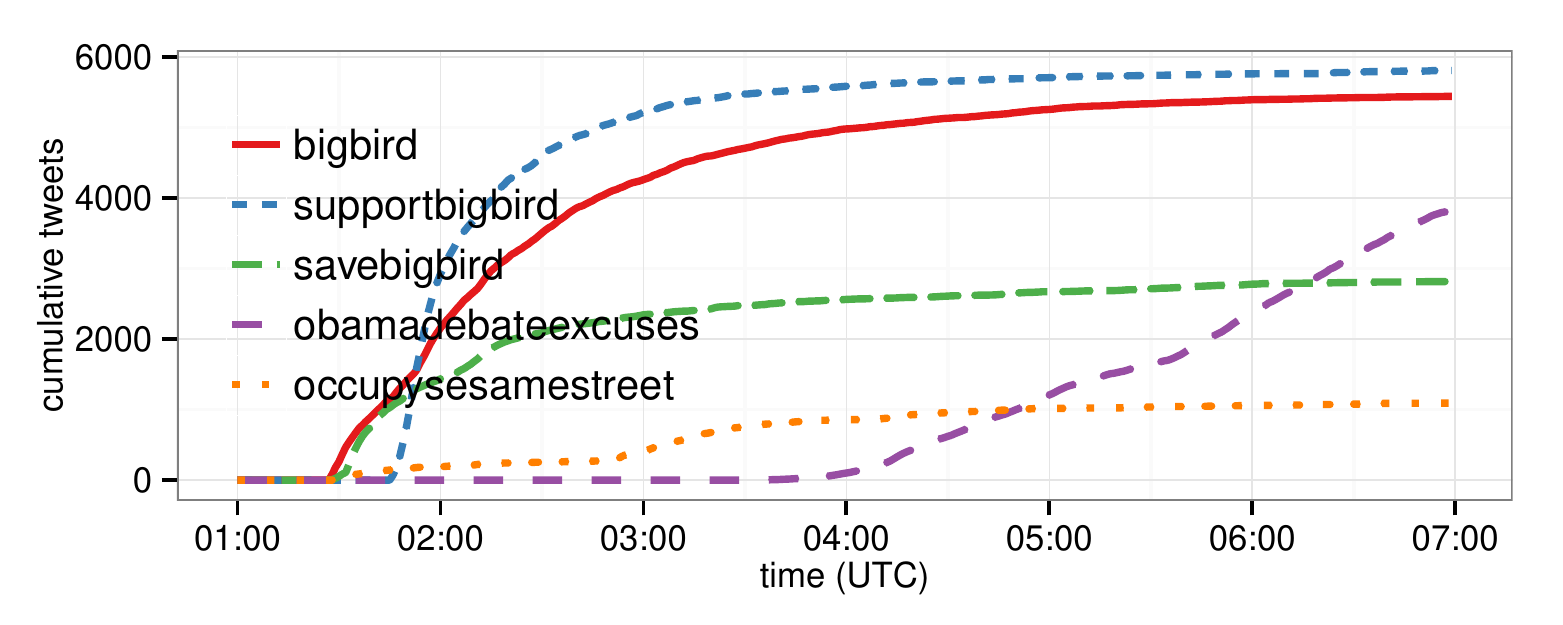} \\
(b) after 48 hours\\
        \includegraphics[trim=.38cm 0cm .38cm 0cm, clip=true,width=1\columnwidth]{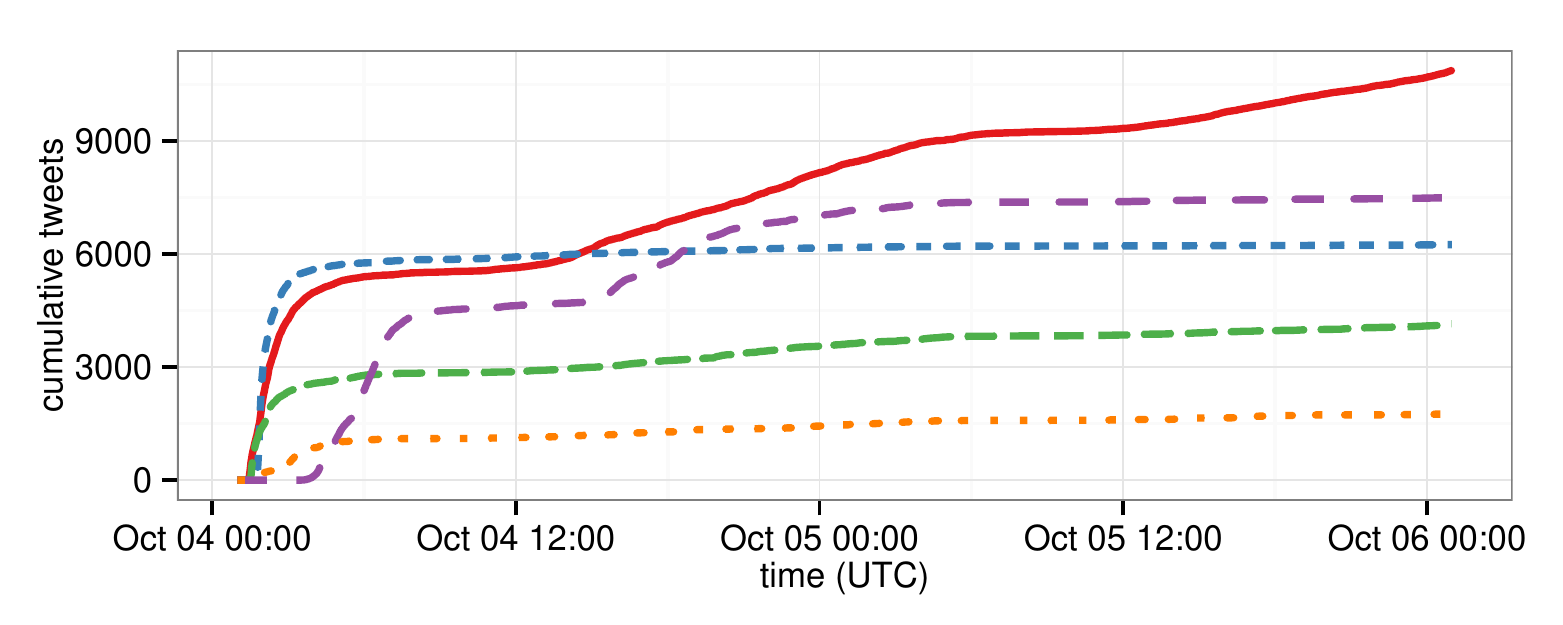} \\
\end{tabular}
\vspace{-1.8em}
    \caption{%
Cumulative tweet volume of top hashtags in the first debate, over the first 6 and 48 hours. Four out of the five hashtags appeared due to Romney's statement regarding PBS, and one was related to the debate more generally.
        \label{fig:debate_6hr}}
\vspace{-.4cm}
\end{figure}

Figure~\ref{fig:debate_3hr}(a) reveals an interesting hashtag dynamics during the first presidential debate: while the hashtag ``\#bigbird'' was born and rapidly grew at around 21:30 (1:30 UTC), the hashtag ``\#supportbigbird'' which was born 15 minutes after took over in about 10 minutes. To see if this is simply the end of the story, we inspect their growth for a longer period. As shown in Figure~\ref{fig:debate_6hr}(a), after six hours from the start of the debate, ``\#supportbigbird'' was still on top of ``\#bigbird.'' However, in a larger time scale as in Figure~\ref{fig:debate_6hr}(b), we discover that ``bigbird'' won back in the $12^{th}$ hour after the debate. The differences between the short-term and long-term dynamics of hashtags lead to an interesting question: How can we characterize the complex dynamics of these emergent hashtags?

From the temporal curves it is clear that the dynamics can not be easily captured by a single process, and hence the parametric modeling approach suggested by prior work (e.g.,~\cite{crane2008robust}) is not appropriate. We focus on analyzing two specific features of the ``pop'' hashtags' adoption dynamics. \textit{Growth} measures the rate of change in hashtag use over observation window. A hashtag that is mentioned 1000 times in 1 hour has higher growth than a hashtag that is mentioned 1000 times in 10 hours.  \textit{Persistence} measures the sustained activity of hashtag use in new tweets over time. A hashtag that continues to be mentioned 70 hours after the debate has greater persistence than a hashtag that stopped being mentioned after 12 hours.

\begin{figure*}[!htb]
\begin{tabular}{ccc}
(a) high growth & (b) high persistence & (c) high growth and high persistence\\
        \includegraphics[trim=.1cm 0cm .5cm 2cm, clip=true,width=.3\textwidth]{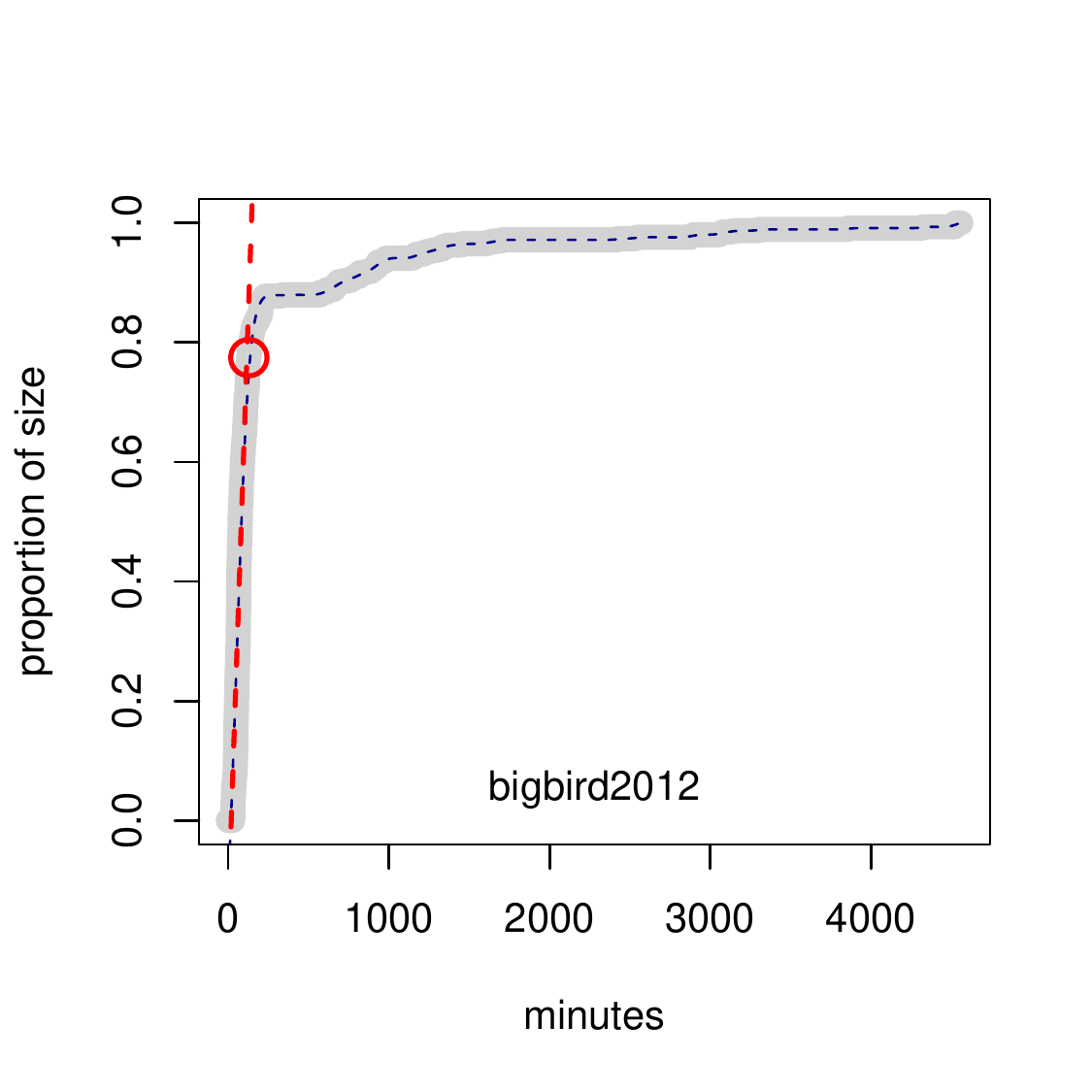} &
        \includegraphics[trim=.1cm 0cm .5cm 2cm, clip=true,width=.3\textwidth]{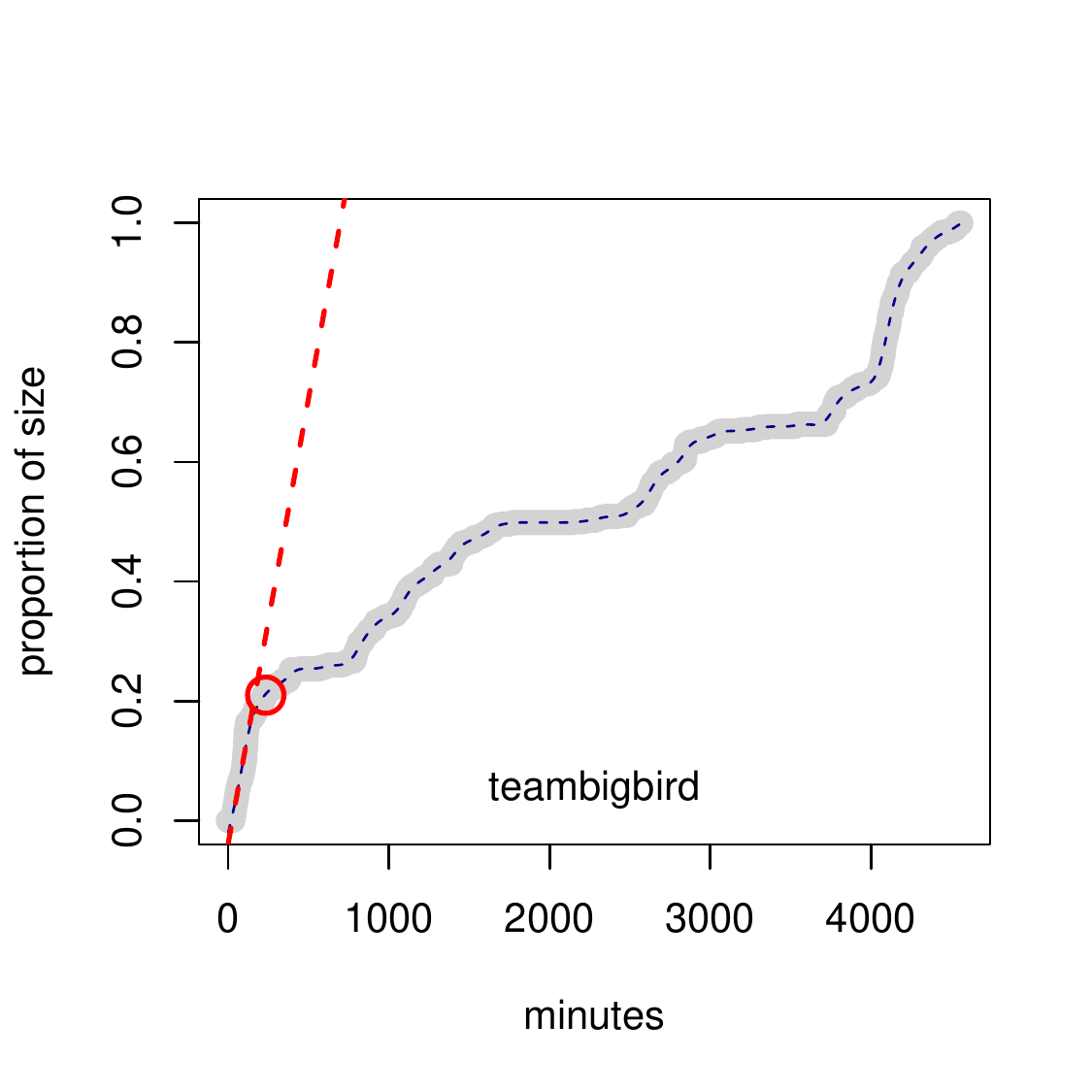} &
        \includegraphics[trim=.1cm 0cm .5cm 2cm, clip=true,width=.3\textwidth]{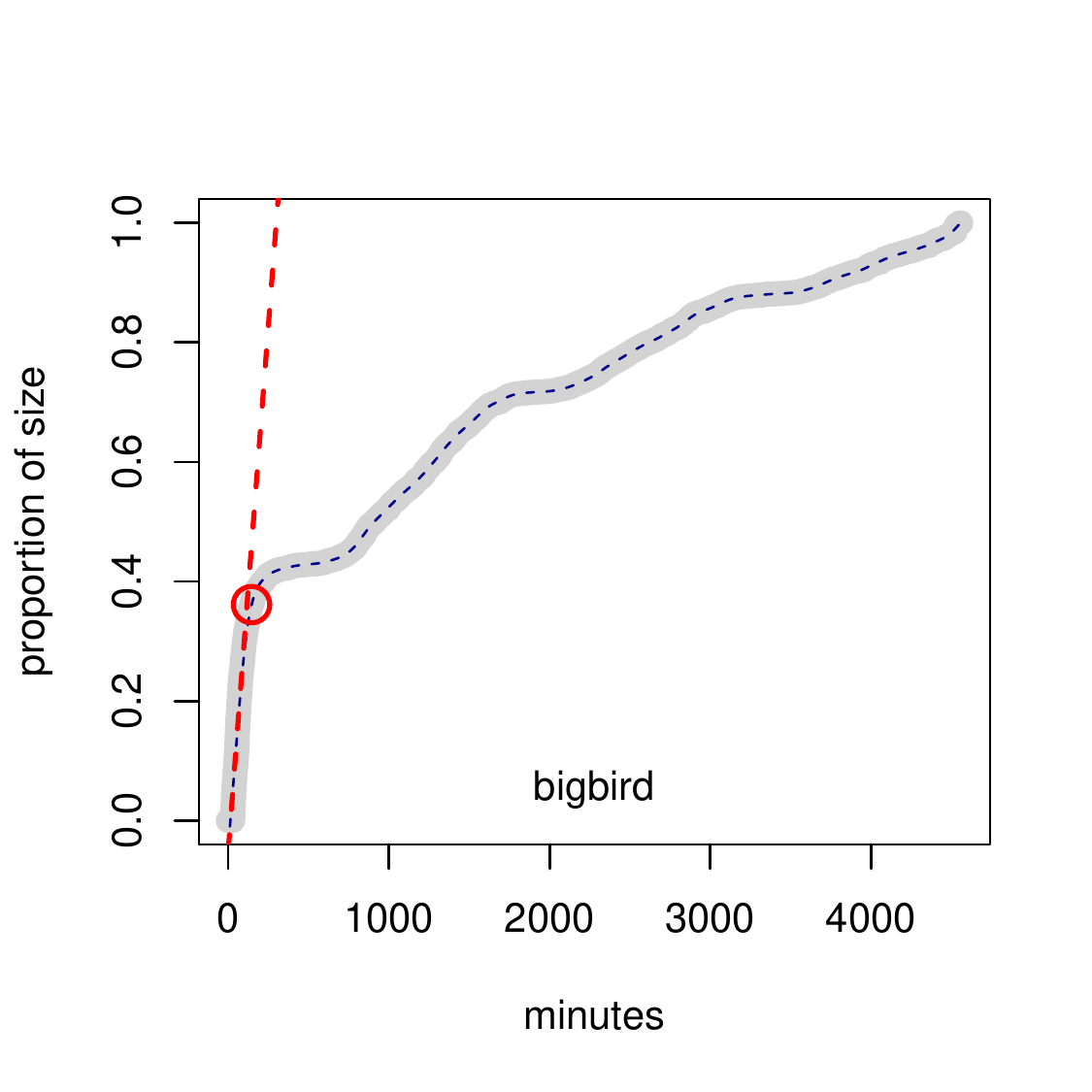} \\
\end{tabular}
\vspace{-2em}
    \caption{%
Characterization of hashtag growth and persistence. We capture the growth of a hashtag as the largest slope (in red line) over its fitted spline function (in blue dashed line).
        \label{fig:curves}}
\vspace{-.6cm}
\end{figure*}

Instead of fitting the curves by a parametric model, we capture the shape of curves by fitting a spline function. Figure~\ref{fig:curves} show three different patterns for the emergent hashtags. A hashtag may grow extremely fast and saturate quickly, as shown in Figure~\ref{fig:curves}(a), grow slower but also slowly saturate, as shown in Figure~\ref{fig:curves}(b), or grow fast and sustain for a longer period, as in Figure~\ref{fig:curves}(c). We then quantify the growth of a hashtag as the largest slope over its fitted spline function. The slope is measured in number of tweets per minute (tpm), so a hashtag with a growth 60 indicates the hashtag gathered 60 tweets per minute in its fast growth phase. 

We identify three critical time points along a hashtag's growth curves:
\begin{itemize}[itemsep=2pt,parsep=0pt,topsep=0pt, partopsep=0pt]
\item \textit{onset time} (\t0): the time where the hashtag was first mentioned in our dataset.
\item \textit{saturated time} (\te): the time where the hashtag's size (in terms of total number of tweets) reach 99\% of its final size in our dataset. We use 99\% instead of final size to avoid the influence from miscellaneous outliers.
\item \textit{turning point} (\tn): the time point where the hashtag growth curve starts deviating from its tangent line of the largest slope. This turning point can be found by a line search procedure along the tangent line. 
\end{itemize}
The persistence is measured as the duration between its onset time \t0 and the saturated time \te. In addition, the turning point divides a hashtag's life-cycle into an initial fast growth phase followed by a relatively slow growing phase before reaching its saturated point.

\section{Results}
We use this population of 2.5 million politically-active Twitter users and 256 ``pop'' hashtags to first describe general patterns of hashtag growth and persistence. Based on behavioral features from the conversational vibrancy framework, we use cluster analysis methods to identify two distinct ``winner'' and ``also-ran'' hashtag classes. Integrating these approaches, we develop statistical models to inductively analyze how features of conversational vibrancy covary for both hashtag classes in the growth and persistence phases of their adoption.

\subsection{Categorizing popular hashtags}

\begin{figure}[!htb]
\begin{tabular}{c}
        \includegraphics[trim=.28cm 0cm .5cm 2cm, clip=true,width=1\columnwidth]{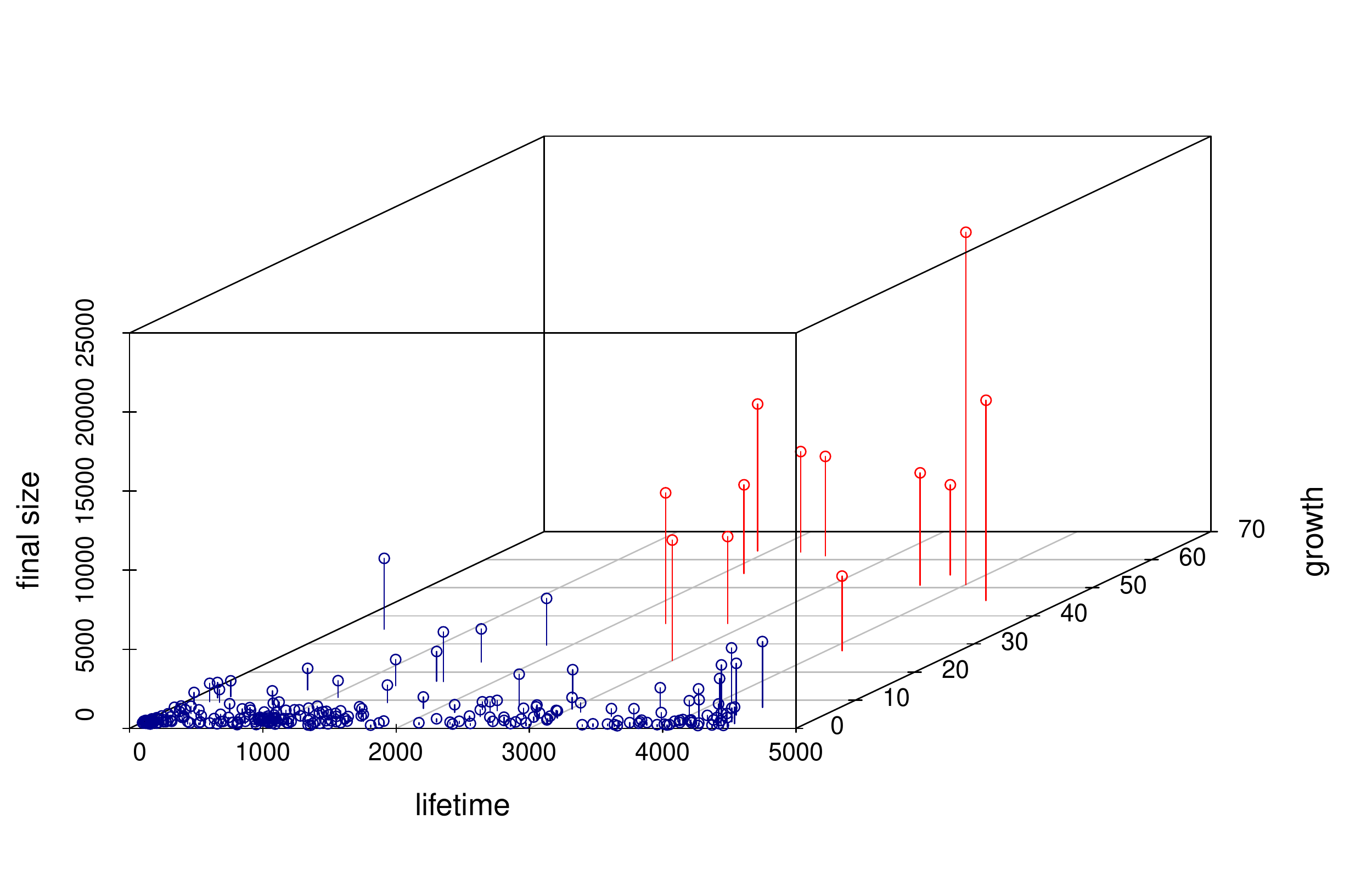} \\
\end{tabular}
\vspace{-3em}
    \caption{%
Two classes of popular hashtags. ``Winners'' (in red) have high growth and persistence (lifetime) with large final size (i.e., total number of tweets). ``Also-rans'' (in blue) are the rest of ``pop'' hashtags.
        \label{fig:cluster}}
\vspace{-.6cm}
\end{figure}

Figure~\ref{fig:cluster} shows the 256 hashtags along the three dimensions: growth, persistence, and final size (the total number of tweets). We use k-means clustering to identify the two distinct classes of hashtags: ``winners'' (in red) grow rapidly and have high levels of persistence while ``also-rans'' (in blue) hashtags have either slower growth or less persistence. The ``winner'' class correspond to the cluster with relatively large final size. The descriptive statistics of the 12 winner hashtags are summarized in Table~\ref{tab:winner_class}. 

\begin{table}[!htb]
    \caption{ Hashtags in the ``winner'' class}
    \label{tab:winner_class}
    \centering\small 
    \begin{tabular}{ clrrr }
    \toprule
DEB&Hashtag & \specialcell{total\\tweets} & \specialcell{growth\\(tpm)} & \specialcell{persistence\\(mins)} \\
    \midrule
1&bigbird & 12667 & 45.45 & 4405 \\
1&obamadebateexcuses&  7617 & 24.12 &2999 \\
1&supportbigbird & 6289 & 61.35 & 2493 \\
1&savebigbird & 4721 & 27.60 & 4118 \\
    \midrule
2&thingsthatmakebidenlaugh & 9290 & 63.10 & 1907 \\
2&malarkey & 7108 & 50.90 & 3669 \\
2&detailsmatter & 6358 & 62.68 & 2249 \\
    \midrule
3&bindersfullofwomen & 22287 & 51.10 & 4003 \\
3&sketchydeal & 5704 & 54.55 & 3733 \\
    \midrule
4&horsesandbayonets & 8266 & 37.27 & 2365 \\
4&strongerwithobama & 5610 & 55.07 & 2162 \\
4&proudofobama & 5502 & 37.28 & 2830 \\
    \bottomrule
    \end{tabular}
\vspace{-.6cm}
\end{table}

\subsection{Explaining Hashtag Growth}
We use time-series regression to discover the relationships between the hashtag growth and different aspects of conversational vibrancy. Our goal is to model the minute-by-minute hashtag growth from the onset up to the turning point. A time-series regression allows us to fit a model of dependent variable (growth) on independent variables (conversational vibrancy) where the serially correlated errors are captured by a linear autoregressive moving-average specification. Concretely, we consider the dependent variable $y_t$ as the number of new tweets for a hashtag at time $t$ where $t$ is between the hashtag onset time \t0 and its turning point \tn.

For each hashtag, we include the following predictors that reflect its associated conversational vibrancy and a control:
\begin{itemize}[itemsep=2pt,parsep=0pt,topsep=0pt, partopsep=0pt]
\item \rt1: the number of new retweets at previous time $t-1$.
\item \rp1: the number of replies at time $t-1$.
\item \src1: the number of new retweet sources up to time $t-1$. 
\item \fol1: the expected largest audience size up to time $t-1$. To compute this, we identify the set of users $U$ who mentioned the hashtag up to time $t-1$, and then compute the expected follower size from the users whose follower size is on the top $10^{th}$ percentile in set $U$.
\end{itemize}

We use the notation $(\cdot)^{\alpha}$ to indicate whether the predictor is an aggregate measure aggregating data from time \t0 to time $t-1$. To fit the linear model, the variables $y_t$, \rt1, \rp1 and \fol1 are log-transformed. We use a Box-Jenkins autoregressive integrative moving average (ARIMA) modeling framework~\cite{durbin2001time} to evaluate the the autocorrelation function and partial autocorrelation function of the residuals, and determine a model with second-order autoregressive process and first-order moving average. The model can be specified as:
\begin{eqnarray*}
y_t &=& \bmat{\beta}^T\mat{x}_{t-1} + \epsilon_t\\
\epsilon_t &=& \phi_1 \epsilon_{t-1} + \phi_2 \epsilon_{t-2} + \nu + \psi \nu_{t-1}
\end{eqnarray*}
where $\mat{x}_{t-1}$ is a vector containing the time-dependent predictors, $\bmat{\beta}$ is a vector of parameters to estimate, $\epsilon$ is the error, $\phi_1$ and $\phi_2$ are the first- and second-order autocorrelation parameters, $\psi$ is the first-order moving-average parameter. The estimated parameters are summarized in Table~\ref{tab:regr_growth}. The table shows parameter estimates (standard errors in parentheses) for the regression models.

In the ``also-ran'' class, hashtag growth has significant and positive associations with \rt1, \rp1 and \fol1. In the ``winner'' class, the growth is positively associated with \rt1 and \fol1, but has a weak and negative associations with \src1. In both classes, hashtags whose tweeters have more followers (\fol1) tend to grow faster. This is consistent with the exposure explanation~\cite{romero2011differences}. Retweets are significant predictors of growth for both winners and also-rans. This finding is consistent with arguments from organizational ecology~\cite{carroll2000demography}, which suggest that organizations and communities with narrow identities that fit closely to the environment tend to thrive in the short term. Hashtags populated by many retweets relevant to the Twitter community at a particular moment appear to grow more quickly, as users are drawn in to the ``hot'' conversation.

The results for replies and growth are not as simple. Replies appear to help also-ran hashtags to grow quickly, but do not appear to help ``winner'' hashtags. For also-ran hashtags, high interactivity would be associated with high growth, as users may be seeking to participate in conversations where it appears others are paying attention to what they say. The fact that replies do not appear to supply resources for the growth of winning hashtags is surprising, but begins to make sense in light of the findings for uniqueness (see below).

The results for uniqueness (\src1) also show a distinction between the two classes. ``Winner'' hashtags appear to be constrained by diversity, whereas ``also-ran''tags are not. Together with the results for replies, this suggests that these classes are distinguished by different conversational vibrancy. ``Winner'' hashtags appear to be those that gain strictly from their relevance to the environment, \textit{i.e.}, they have interesting tweets. When fitness is high, diversity becomes a drag~\cite{kauffman1993origins}. 
By contrast, ``also-ran'' tags appear to be less relevant. 
At the same time, what is important for growth is the ability to bring something beyond relevance---interaction with others. 



\begin{table}[!tb]
    \caption{ Growth models}
    \label{tab:regr_growth}
    \centering\small 
    \begin{tabular}{ lll }
    \toprule
Variables&Winner&Also-ran\\
    \midrule
\rt1&0.0626** (0.0239)&0.2651*** (0.0073)\\
\rp1&-0.0128 (0.0161)&0.1642*** (0.0108)\\
\src1&-0.0016*** (0.0002)&0.0000 (0.0000)\\
\fol1&0.1048** (0.0375)&0.0891*** (0.0042)\\
    \midrule
Loglik&-476.54&-26760.69\\
AIC&971.07&53539.38\\
    \bottomrule
    \end{tabular}
\vspace{-.6cm}
\end{table}

\subsection{Explaining Hashtag Persistence}
To examine the effect of earlier conversational vibrancy on a hashtag's persistence, we use survival analyses based on the Cox proportional-hazards model~\cite{cox1972regression}. We examine the survival time of a hashtag starting from the turning point \tn to the saturated time \te. Let $T$ be a continuous random variable, the survival function is defined as: $S(t) = Pr(T > t)$, the probability that the hashtag will survive (i.e., not saturated) beyond $t$.

A hazard function assesses the instantaneous risk of demise at time $t$, conditional on survival to that time. 
\[
h(t) = lim_{\Delta t\rightarrow 0}\frac{Pr[(t\leq T < t+\Delta t)]}{\Delta t}
\]
In the Cox regression model, the log hazard can be specified as:
\[
h(t) = h_0(t)exp(\bmat{\beta}^T\mat{x})
\]
where $\mat{x}$ is a vector of predictors, $\bmat{\beta}$ is a vector of parameters to estimate, $h_0(t)$ is the baseline hazard.

To capture a hashtag's earlier conversational vibrancy, we focus on activities occurring in its fast growth phase. Hence the predictors we choose to examine are aggregate measures for each hashtag at the turning point \tn, including:
\begin{itemize}[itemsep=2pt,parsep=0pt,topsep=0pt, partopsep=0pt]
\item \rtT: the total number of retweets received up to the turning time point \tn, i.e., between \t0 and \tn.
\item \rpT: the total number of replies posted up to time \tn.
\item \srcT: the total number of retweet sources up to time \tn.
\item \folT: the expected largest audience size up to time \tn. The computation is similar to \fol1 except that the time is fixed at \tn.
\end{itemize}

The notation $(\cdot)^{\alpha}$ indicates that the predictor is an aggregate measure. Table~\ref{tab:regr_lifetime} summarizes the results for the survival analyses. All coefficient are exponentiated and thus interpretable as multiplicative effects on the hazard (standard error in parenthesis are not exponentiated). For the winner class, the predictors \rtT~and \rpT~have significant coefficients. For example, an additional reply reduces the per minute hazard of hashtag saturation by 0.65\%. A positive coefficient increases the value of the hazard function and therefore indicates a negative effect on survival time. In other words, \rpT~has a positive association with a hashtag's persistence and \rtT~has a negative association with the persistence. The other predictors, \srcT~and \folT, have no significant effect. For the also-ran class, \rtT~has negative effect while \srcT~has positive effect on the hashtag persistence. 

As expected, and consistent with exposure explanations, the number of followers is not a significant predictor of a tag's longevity. Also consistent with the explanation for growth is the finding for retweets. 
Organizational ecology predict that organizations and communities with specific identities strongly coupled with the environment are risk of demise when the environment changes~\cite{carroll2000demography}. The very thing that made them appealing when the environment was suited to what they had to offer now makes them less useful. For example, a hashtag that mainly contains jokes about the Big Bird reference may be popular in the initial frenzy after the comment. As the debate moves on or ends and serious issues come to the fore, the humorous hashtags lose their relevance to the new environment.

\begin{figure}[!tb]
\hspace*{-.8cm}
\begin{tabular}{cc}
(a) Winner&(b) Also-ran\\
        \includegraphics[trim=.1cm 0cm 1cm 2cm, clip=true,width=.49\columnwidth]{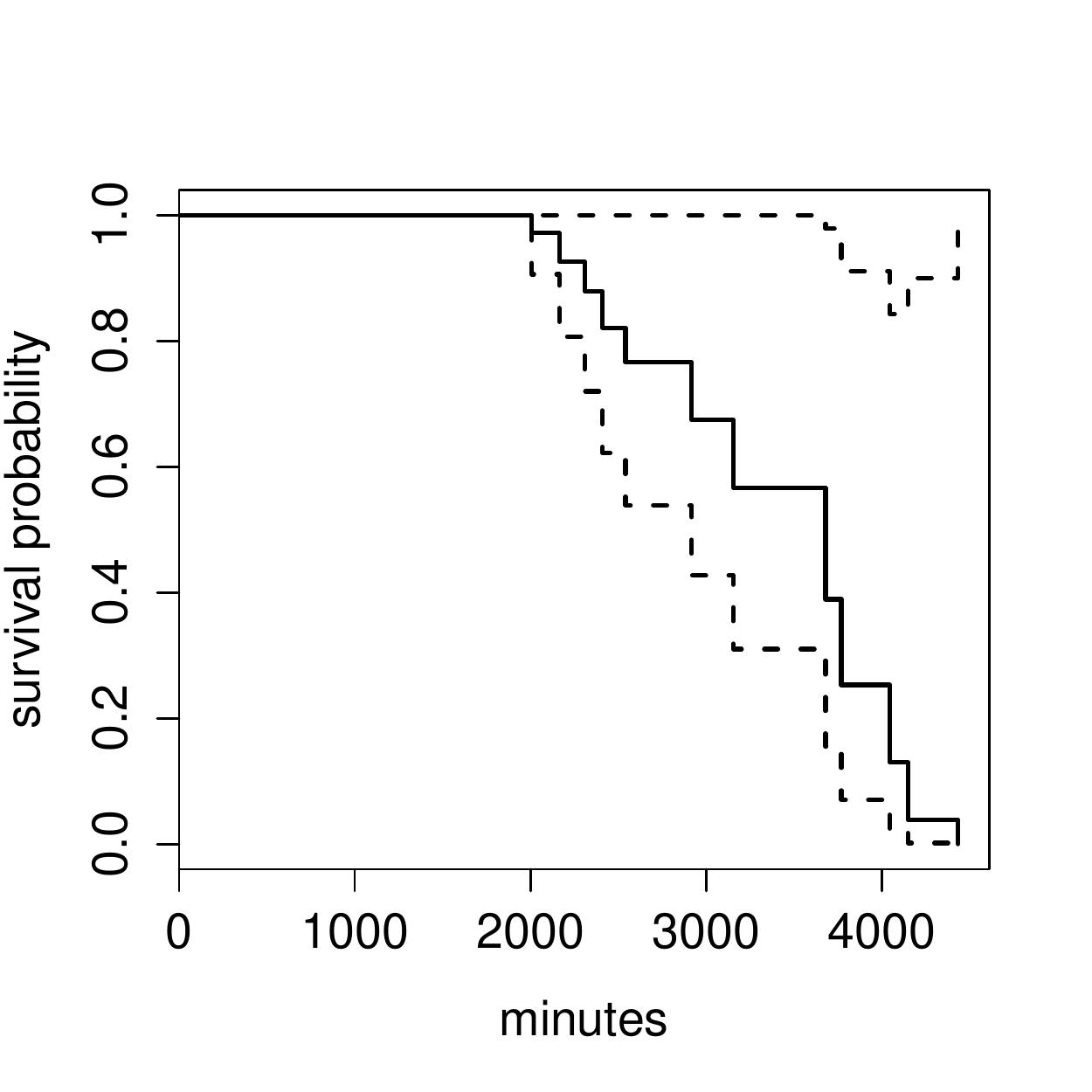} &

        \includegraphics[trim=.1cm 0cm 1cm 2cm, clip=true,width=.49\columnwidth]{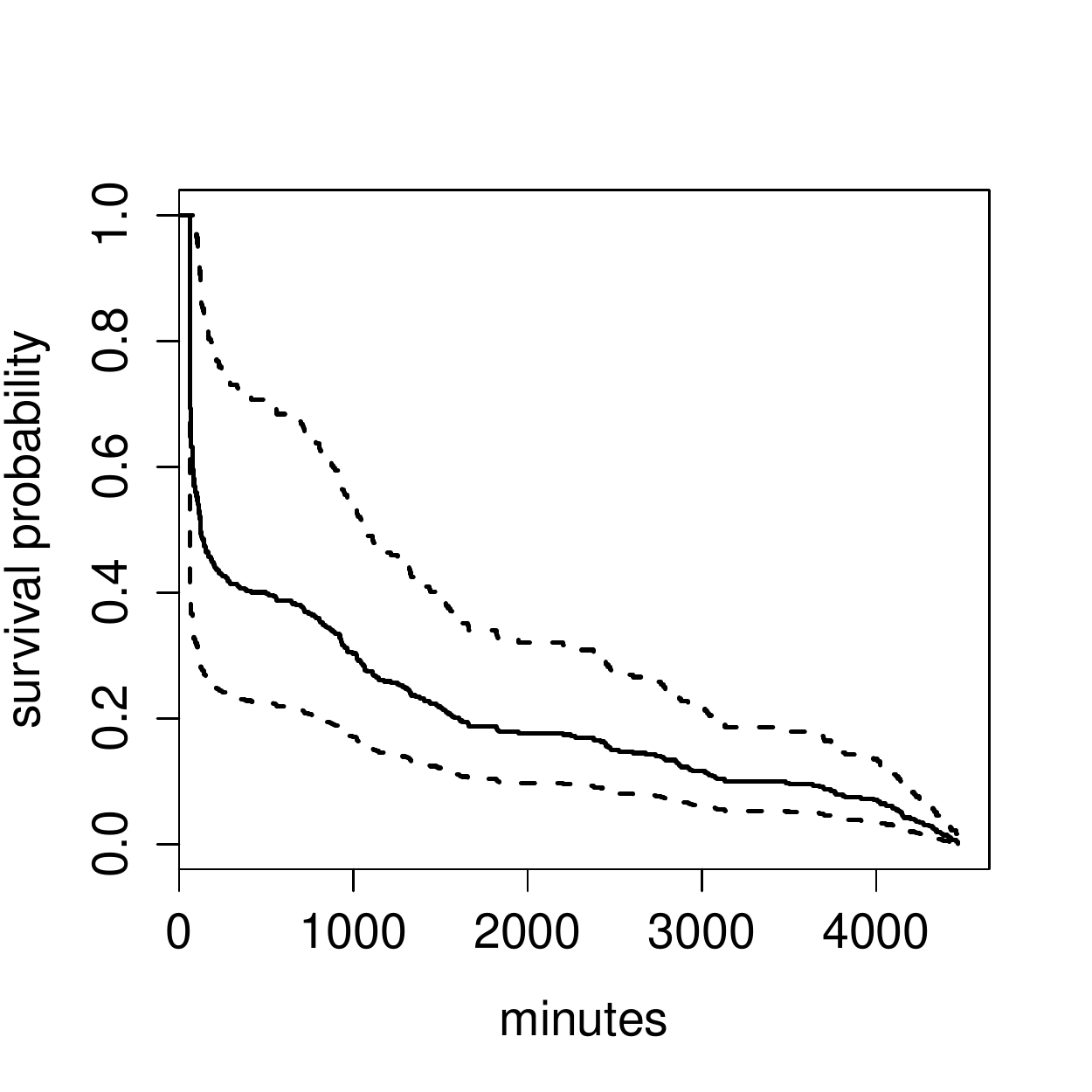} \\
\end{tabular}
\vspace{-2em}
    \caption{%
Estimated survival function $\hat{S}(t)$ for the two classes. Half of the hashtags in the winner class lived beyond 2.5 days, and half of the hashtags in the also-ran class died out (become saturated) in two hours. The dash lines show a point-wise 95\% confidence envelope around the survival function.
        \label{fig:survival}}
\vspace{-.5cm}
\end{figure}

Interactivity and diversity present a more complex set of findings. More replies increase the longevity of ``winner'' hashtags while having no effect for ``also-rans.'' Simultaneously, diversity extends the life of also-rans but has no effect for winners. This is additional evidence of a qualitative difference between the two classes of hashtags. Replies appear to buffer winner tags from the inevitable decline in their relevance over time \cite{baum1991institutional}. That is, ``winner'' hashtags with many replies may able to fall back into the niches that were originally occupied by the ``also-rans''---a place to exchange ideas about a topic, rather than just to broadcast one's opinions or humorous insights. The fact that ``also-rans'' benefit from diversity, but ``winners'' do not, suggests that some ``also-rans'' may build up a sufficient community that wishes to continue the conversation even as new topics emerge and relevance declines. Fig.~\ref{fig:survival} shows the estimated survival function $\hat{S}(t)$ for the two classes. In the winner class, half of the hashtags lived beyond 2.5 days. In the also-ran class, half of the hashtags died out (become saturated) in two hours. 

\begin{table}[!tb]
    \caption{Persistence models}
    \label{tab:regr_lifetime}
    \centering\small 
    \begin{tabular}{ lll }
    \toprule
Variables&Winner&Also-ran\\
    \midrule
\rtT&1.001* (0.0003)&1.001* (0.0003)\\
\rpT&0.9935* (0.0003)&1.000 (0.0002)\\
\srcT&1.001 (0.0014)&0.9933*** (0.0019)\\
\folT&1.000 (0.0000)&1.000 (0.0000)\\
    \midrule
Loglik&-15.92968&-1028.731\\
AIC&39.85936&2065.461\\
    \bottomrule
    \end{tabular}
\vspace{-.4cm}
\end{table}

\subsection{Examining Environmental Context}
In our analysis, all the ``pop'' hashtags emerged from a unusual condition where users' attention was concentrated on debate events. It is possible that a hashtag's growth and death is largely determined by this particular environmental condition rather than the conversational vibrancy of the hashtag itself. To test the robustness of our findings against these concerns, we re-construct the prior growth and persistence models and include variables for users' tweeting activity for other hashtags at the same moment.

We extend the growth models to include three time-dependent environmental covariates. For each hashtag $i$, we have:
\begin{itemize}[itemsep=2pt,parsep=0pt,topsep=0pt, partopsep=0pt]
\item \rtE1: the total number of new retweets that do not contain hashtag $i$, at previous time $t-1$.
\item \rpE1: the number of replies that do not contain hashtag $i$, at time $t-1$.
\item \srcE1: the number of new retweet sources that do not contain hashtag $i$, up to time $t-1$.
\end{itemize}

Based on the time-series regression described above, we include the three new predictors into new models of growth. The estimated results are summarized in Table~\ref{tab:regr_growth2}. Interestingly, none of the environmental covariates has significant effect on the growth of winner hashtags. However, the predictors \rtE1 and \rpE1 are positively correlated to the growth of ``also-ran' hashtags. While these effect are not as strong as \rt1 and \rp1, the results suggest that ``also-ran'' hashtags benefit from other tweeting activities.

\begin{table}[!tb]
    \caption{ Growth models with co-occurring hashtags}
    \label{tab:regr_growth2}
    \centering\small 
    \begin{tabular}{ lll }
    \toprule
Variables&Winner&Also-ran\\
    \midrule
\rt1&0.0596* (0.0240)&0.2817*** (0.0073)\\
\rp1&-0.0141 (0.0161)&0.1560*** (0.0109)\\
\src1&-0.0016*** (0.0002)&0.0013*** (0.0003)\\
\fol1&0.0989** (0.0380)&0.0811*** (0.0042)\\
    \midrule
\rtE1&-0.0028 (0.0271)&0.0530*** (0.0043)\\
\rpE1&0.0113 (0.0174)&0.0173*** (0.0038)\\
\srcE1&0.0014 (0.0016)&0.0000 (0.0000)\\
    \midrule
Loglik&-475.89&-26793.36\\
AIC&975.79&53610.72\\
    \bottomrule
    \end{tabular}
\vspace{-.4cm}
\end{table}

\begin{table}[!tb]
    \caption{ Persistence models with co-occurring hashtags}
    \label{tab:regr_lifetime2}
    \centering\small 
    \begin{tabular}{ lll }
    \toprule
Variables&Winner&Also-ran\\
    \midrule
\rtT&1.001 (0.0004)&1.001** (0.0003)\\
\rpT&0.9904* (0.0045)&1.000 (0.0002)\\
\srcT&1.002 (0.0021)&0.9928*** (0.0019)\\
\folT&1.000 (0.0000)&1.000 (0.0000)\\
    \midrule
\rtEt&1.428 (2.688)&0.7173* (0.1502)\\
\rpEt&1.548 (2.288)&0.6976* (0.1709)\\
\srcEt&2.023 (1.560)&1.020 (0.01799)\\
    \midrule
Loglik&-11.65331&-1023.476\\
AIC&37.30663&2060.952\\
    \bottomrule
    \end{tabular}
\vspace{-.4cm}
\end{table}

To study the relationship between hashtag's persistence and the environment, we include three new predictors in the survival analyses. Unlike the original four predictors in the persistence models which were measured at a fixed point in time (\tn), the new predictors are measured at each time $t$, where $t$ is between the turning point \tn~and the saturated time \te. The time-dependent covariates measures ``what was going on in the environment'' before the hashtag's saturation. For each hashtag $i$, we have:
\begin{itemize}[itemsep=2pt,parsep=0pt,topsep=0pt, partopsep=0pt]
\item \rtEt: the total number of new retweets that do not contain hashtag $i$, at time $t$.
\item \rpEt: the number of replies that do not contain hashtag $i$, at time $t$.
\item \srcEt: the number of new retweet sources that do not contain hashtag $i$ up to time $t$.
\end{itemize}

These three new time-dependent covariates, together with the four fixed point covariates, are included in the new persistence models. The estimated results are summarized in Table~\ref{tab:regr_lifetime2}. Like the growth model, we see no impact from the environment on this model for predicting winner hashtags' persistence. For also-ran hashtags, the \rtEt~and \rpEt~increase the hazard of becoming saturated and they are negatively associated with persistence. The reflects the fact that many of the also-ran hashtags died out in a couple of hours where there were still high tweeting activities going on in the environment. 

\section{Discussion}
User-generated content with high levels of conversational vibrancy should be a focal point for attracting the attention and participation of users. These elements should satisfy a confluence of users' needs by providing topical information, interactive sociality, diverse and novel content, and involvement from well-connected users and lead to a feedback loop that reinforces these tendencies. Operationalizing this model for Twitter in particular, we investigated how these features of conversational vitality influenced both the growth and persistence of adoption for different classes of hashtags.

\textbf{Summary of findings.} The number of times a hashtag is retweeted (topicality) as well as the popularity of the users mentioning the hashtag (prominence) lead to more rapid growth of hashtags for ``winner'' and ``also-ran'' hashtags alike. Additional replies (interactivity) and the number unique retweet sources (diversity) support the persistence of ``winner'' and ``also-ran'' hashtags, respectively. In addition, our findings unexpectedly suggest that the number retweets inhibit the growth of hashtags. Because these findings remain robust after controlling for other contextual behavior of users, these findings provide mixed evidence for the simple cumulative vibrancy model we proposed. In particular, the fact that additional incremental activity inhibits the growth and persistence of hashtags suggests there higher-order processes that lead to limits or tipping points.

One direction for extending this conversational vibrancy model would be to draw upon theories from organizational ecology that describe the birth, growth, and death of organizations and communities.  Ecologists treat organizations as entities which, like hashtags, coordinate behavior through identities that suggest the kinds of behaviors and messages that are appropriate within organizations of a particular form~\cite{carroll2000demography}. In particular, ecological research focuses on the importance of balancing consistency and coherence with flexibility and access to resources. The environment provides a limited supply of resources for organizations (laborers, customers, \textit{etc.}) analogous to how the Twitter environment provides a limited supply of resources for hashtags (attention, users, ideas to express, \textit{etc.}).  Organizations survive and thrive when their identities are specific enough such that individuals know what to expect from them but broad enough that they can address a range of needs and access a variety of resources. It has been recognized that organizations with similar identities tend to thrive as their population grows, with each gaining more attention and legitimacy. At some point a limit is reached and the environment can no longer support most of these organizations, and only a few remain~\cite{carroll2000demography}.

A substantial portion of the research on growth and sustainability involves the comparison of cultural forms, such as words, names, memes, or networks.  Both exposure and ``fitness'' play important roles in these models, suggesting rationales for both self-reinforcing growth and saturation ~\cite{steyvers2010large,barabasi2003linked}. Studies of baby naming conventions have found that burstiness is associated with short life-time because collective aversion to ``faddishness'' limits the growth of a name~\cite{berger2009adoption}. These findings are corroborated by other work on memes that suggest social media have carrying capacities~\cite{leskovec2009meme}. Our model introduces an additional factor to consider: the endogenous and developing qualities of the communication and interactions between those that use the form, serving to weaken the determinism suggested by fitness and first-mover based explanations.

The application of this model to linguistic evolution may be particularly appealing.  The ways in which linguistic conventions are shared by a groups have been investigated in the field of language dynamics~\cite{loreto2011statistical}. Computational models have shown that simple linguistic interactions can lead the group to reach a consensus on a given linguistic convention by the progressive elimination of competing synonyms~\cite{steels1995self,baronchelli2006sharp}. Interestingly, these models assume that the competing synonyms are perfectly equivalent. While the situation is clearly more complex with hashtags, our research demonstrates that the observed winner-takes-all dynamics are not necessarily driven only by intrinsic properties of hashtags but also by other contextual and structural features which suggest new avenues for research across fields like communication, social computing, linguistics, and political science.

Our approach in this analysis also has limitations that merit future research. We do not examine more platform-specific contextual features such as the effect of a hashtag's length (number of characters) on its life-cycle, which has significance due to Twitter's 140-character limit for a message~\cite{yang2012we}. Despite the highly situational and socially-embedded interactions in social emdia, our modeling approach assumed the observation of hashtags on users are the result of an independent process. Future work could specify alternative models that account for the local clustering of ties or the modular structure of these communities which likely strongly influence the growth and persistence of hashtag use. 

\section{Conclusion}
We introduced a theoretical framework called conversational vibrancy that synthesizes work from social computing and mass communication to understand why some forms of user-generated content are adopted while others are abandoned. The U.S. presidential debates allowed us to understand the lifecycle of hashtag adoption and use in response to exogenous shocks that allow us to separate out the confounding influence of topical relevance and social exposure. Using simple computational means to characterize the key aspects of hashtags's complex dynamics, statistical models reveal qualitative differences in how conversational vibrancy influences the growth and persistence of distinct classes of hashtags. The study also demonstrates the power of computational social science~\cite{lazer2009social} approaches to meld data-driven computational and statistical approaches to investigate fundamental questions about human sociality.

\section{Acknowledgments}
The authors gratefully acknowledge the support of the Lazer Lab at Northeastern University, supported in part by MURI grant \#504026, DTRA grant \#509475, and ARO \#504033.

\bibliographystyle{plain} \footnotesize
\bibliography{bigbird_arxiv}

\end{document}